\documentclass[traditabstract]{aa} 
\usepackage{amstext}
\usepackage{amsmath,bbold,mathrsfs,bm}
\usepackage{graphicx}
\usepackage{textcomp}
\usepackage{txfonts}
\usepackage{time}
\usepackage{subfig}
\usepackage{color}
\usepackage{url}
\usepackage{upgreek}

\usepackage{natbib}
\bibpunct{(}{)}{;}{a}{}{,}

\usepackage{url}
\usepackage{epic}
\usepackage{rotating}
\usepackage{float}

\usepackage{multirow} 
\usepackage{array} 
\usepackage{balance}
\usepackage{epic}
\usepackage{xcolor}
\usepackage[breaklinks=true]{hyperref} 

\usepackage{xifthen}

\makeatletter
\def\url@leostyle{%
  \@ifundefined{selectfont}{\def\UrlFont{\sf}}{\def\UrlFont{\tiny\ttfamily}}}
\makeatother
\begin{document}

\title{CRISPRED: A data pipeline for the CRISP imaging spectropolarimeter}

\author{J. de la Cruz Rodr\'iguez\inst{1,2} \and M.G.
  L{\"o}fdahl\inst{1} \and P. S\"utterlin\inst{1} \and
  T. Hillberg\inst{1} \and L. Rouppe van der Voort\inst{3}}  

\institute{Institute for Solar Physics, Dept. of Astronomy, Stockholm
  University, Albanova University Center, 106\,91 Stockholm, Sweden
  \and Department of Physics and Astronomy, Uppsala University, Box
  516, 751\,20, Uppsala, Sweden \and Institute of
  Theoretical Astrophysics, University of Oslo, P.O. Box 1029
  Blindern, N--0315 Oslo, Norway}

\date{Revised 2014-09-24} 

\frenchspacing 

\abstract{The production of science-ready data from major solar
  telescopes requires expertise beyond that of the typical observer.
  This is a consequence of the increasing complexity of instruments
  and observing sequences, which require calibrations and corrections
  for instrumental and seeing effects that are not only
  difficult to measure, but are also coupled in ways that require
  careful analysis in the design of the correction procedures. Modern
  space-based telescopes have data-processing pipelines capable of
  routinely producing well-characterized data products.
  High-resolution imaging spectropolarimeters at ground-based
  telescopes need similar data pipelines.
%
%
  { We present new methods for flat-fielding spectropolarimetric
    data acquired with telecentric Fabry-Perot instruments and a new
    approach for accurate camera co-alignment for image restoration. We}
  document a procedure that forms the
  basis of current state of the art processing of data from the CRISP
  imaging spectropolarimeter at the Swedish 1-m Solar Telescope (SST).
  By collecting, implementing, and testing a suite of computer
  programs, we have defined a data reduction pipeline for this
  instrument. 
 This pipeline, CRISPRED, streamlines the process of
  making science-ready data. 
 { It} is implemented and operated in IDL, with time-consuming
  steps delegated to C. { CRISPRED} will be the basis also for the
  data pipeline of the forthcoming CHROMIS instrument.}

 


\maketitle

\section{Introduction}
\label{sec:introduction}

During the past 20 years, a list of Fabry-P\'erot interferometers
(FPI) have been developed and installed in many ground based
facilities and airborne missions, e.g: TESOS at the VTT
\citep{1998thesos,2002SoPh..211...17T}, IBIS at the Dunn Solar
Telescope \citep{2006cavallini}, CRISP at the Swedish 1-m Solar
Telescope \citep{scharmer06comments}, IMAX in the Sunrise mission
\citep{2011SoPh..268...57M}, GFPI at Gregor
\citep{2013OptEn..52h1606P} and many more that are forthcoming. These
instruments allow acquisition of spectropolarimetric data of a
relatively large field-of-view (FOV) with very high cadence, very high
spatial resolution, and an acceptable wavelength coverage
\citep[e.g.,][among
others]{2004A&A...415..717T,2008A&A...480..515C,2008A&A...480..265B,scharmer08crisp,
  2010ApJ...723L.180S,2012ApJ...752L..12D}.

In principle, FPIs can be mounted in telecentric (TESOS, CRISP) or
collimated setup (IBIS, GFPI, IMAX). Most difficulties related to the
processing of data from FPI instruments arise from microscopic
corrugations in the surface of the etalons. In telecentric
configuration, these aberrations are close to the focal plane, and
therefore, the resulting data suffer from \emph{undesired} wavelength
shifts across the FOV. The unavoidable presence of atmospheric
distortions from the ground further complicates the problem by
introducing differential motions and aberrations in the image: between
acquisitions, solar features are moved and distorted on the FOV, so
there is not a unique wavelength shift (or pixel) that can be
associated to that feature as it may have sampled a range of pixels in
the different exposures and wavelengths. Image restoration techniques,
currently in widespread use for solar data, are not designed to deal
with such wavelength inhomogeneities, and often one has to resort to
using approximations that make the images \emph{smoother} prior to
image restoration \citep[e.g.,][]{2011schnerr}. In this paper, we
propose a new method to accurately perform image restoration on
datasets acquired with FPIs in telecentric setup.

Furthermore, a recurring problem inherent to most instruments
mentioned above, has been the lack of well established methods and
techniques to properly reduce and process the resulting datasets,
although some efforts have been made
\citep[e.g.,][]{2008A&A...481..897R,2011schnerr}, and certain
corrections are standard to all instruments: dark current and
flat-field correction, pre-filter compensation and, in most cases,
image restoration.

The task of going from raw data to science ready data is beyond the
skills of the majority of potential users due to the complexity of the
data calibration and required compensation for seeing degradation.
There is thus a strong incentive to relieve solar scientists from the
burden of gaining expert knowledge about state of the art
instrumentation by supplying them with easy to use data pipelines with
well-characterized data products. Well-designed data pipelines allow
such instrumentation to reach their full scientific potential, while
minimizing the efforts of the user and opening up access to the
facility to inexperienced users. Such pipelines will also be necessary
in the future when solar telescopes are routinely operated in service
mode and are required to deliver science-ready data to scientists
\citep{2009ASPC..415..332R}. The European solar community has
recognized the need for such pipelines and therefore included pipeline
development in the recently started EU-funded SOLARNET project.

In this paper, we {also} describe
CRISPRED\footnote{\url{http://dubshen.astro.su.se/wiki/index.php?title=CRISPRED}},
a data reduction pipeline for CRISP. The pipeline started as a
collection of scripts written by different (present as well as former)
researchers at the Institute for Solar Physics (ISP), Uppsala
University, and University of Oslo, coded in a variety of languages.
These heterogeneous pieces of code have since been incorporated into
an object-oriented IDL structure with computationally demanding parts
performed in C subprograms called as dynamically loadable modules
(DLMs). Early versions of CRISPRED have been used by several authors
to reduce and prepare their data, given that a basic framework has
been functional since 2011
\citep[e.g.,][]{2012watanabe,2013scharmer,2013delacruz,2013ApJ...774...32V,%
  2013ApJ...769...44S,2013ApJ...774..123W,2013rouppe,2014leenaarts}.

A pipeline for an instrument like CRISP must be based on a detailed
knowledge about the formation of the image data, including the various
sources of measurement error and the calibrations that allow them to
be characterized. Section \ref{sec:image-form} describes the image
formation relevant for FPI-based imaging spectropolarimeters like
CRISP.

The calibration and removal of the errors and contaminations are
important parts of the pipeline. Newly developed methods are
introduced in Sect.~\ref{sec:new-developments}, before we describe the
CRISPRED pipeline in Section \ref{sec:data-reduction-steps}. We end
with a discussion on future developments in
Sect.~\ref{sec:discussion}.

\section{Image formation and optical setup}
\label{sec:image-form}

The signal that we want to detect with an imaging spectropolarimeter,
intensity and polarization of visible light, is formed in the Sun's
photosphere and chromosphere. The intensity pattern over the FOV is
modulated by polarization optics and the final image is the
convolution of this intensity pattern with the telescope PSF at the
wavelength selected by narrowband (NB) filters.

The image formation is quite complex. Turbulence in the Earth's
atmosphere degrades the image quality and causes spectral cross-talk.
The telescope and the optical setup introduce undesirable
polarization. The filter profiles vary over the FOV and introduce
other unwanted effects. The following sections describe the image
formation, including an overview of the most important deviations from
the ideal case.

The optical setup and the details of the instrument are obviously
important for the formation of the images. The CRISPRED pipeline is
developed for SST/CRISP, so we provide a description of that
particular instrument for the following discussion. We believe most of
the CRISPRED methods described in later sections are easily adaptable
to similar instruments but we will point out when a certain feature in
the setup is necessary. More details of the CRISP setup can be found
in the appendix.

\subsection{Atmosphere and telescope}
\label{sec:atmosphere-telescope}

Turbulence randomly mixes air with varying temperature (and therefore
refractive index), causing perturbations in the wavefront phases that
vary spatially with a time scale of ms. We refer the interested reader
to \citet{1965JOSA...55.1427F} and \citet{roddier99theoretical} for
mathematical descriptions of these effects. Turbulence near the
telescope aperture causes global motion and blurring of the image.
Similar effects at higher altitudes ($\sim$10~km) take the form of
variations over the FOV, where the seeing causes rubber-sheet
deformations of the image and blurring that varies over small scales.
Scattering from aerosols can produce a diffuse component that lowers
the contrast. Particularly in \emph{calima} conditions, when the
Saharan air layer, an elevated (1000--5500~m) layer of dry air and
mineral dust originating from the Saharan desert, forms over large
parts of the northern equatorial Atlantic
\citep[e.g.,][]{2004BAMS...85..353D}. The polarizing effects from the
atmosphere are visible in the blue sky but are small in direct
sunlight.

\begin{figure}[tb]
  \centering
  \includegraphics[bb=135 109 549 671,clip,width=\linewidth]{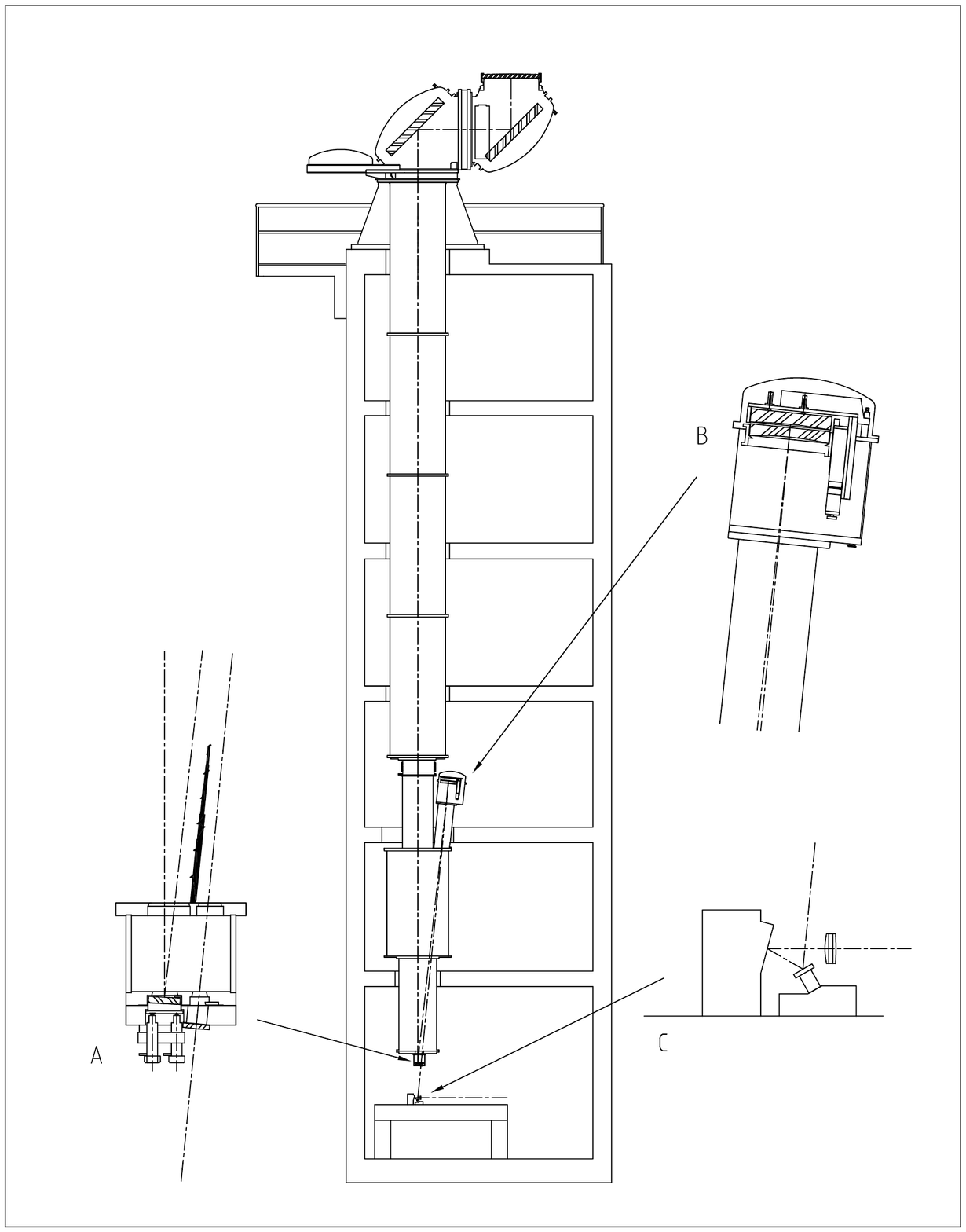}
  \caption{Sketch of the optical path of SST, from the 1-m lens, via
    the two alt-az mirrors, the field mirror on the bottom plate
    (inset A), the Schupmann corrector (inset B), the field lens and
    exit window (inset A), to the tip-tilt mirror, deformable mirror,
    and re-imaging lens on the optical table (inset C). The optical
    path continues in Fig.~\ref{fig:opticalpath}. Courtesy of G\"oran
    Scharmer.}
  \label{fig:sst}
\end{figure}

The SST is located on the island of La Palma, a little less than
2400~m above sea level in a 17-m high tower. Figure~\ref{fig:sst}
shows the optics of the telescope. Via the two 45\degr\ turret alt-az
mirrors, the beam is directed down through the vertical telescope tube
to a primary focus on the bottom plate. The alt-az configuration
causes image rotation and the rotation of the lens and two mirrors
cause time-varying polarization \citep{selbing05sst}, mostly far from
focus. The lens itself is a singlet and therefore has enormous
chromatic aberrations. These are removed by a Schupmann corrector,
that is fed from a field mirror on the bottom plate of the tube and
forms a new focus just outside an exit vacuum window behind a large
baffle.

\subsection{Pre-CRISP optics}
\label{sec:pre-crisp-optics}

\begin{figure}[tbp]
  \centering
  \includegraphics[bb=134 448 449 687,clip,width=\linewidth]{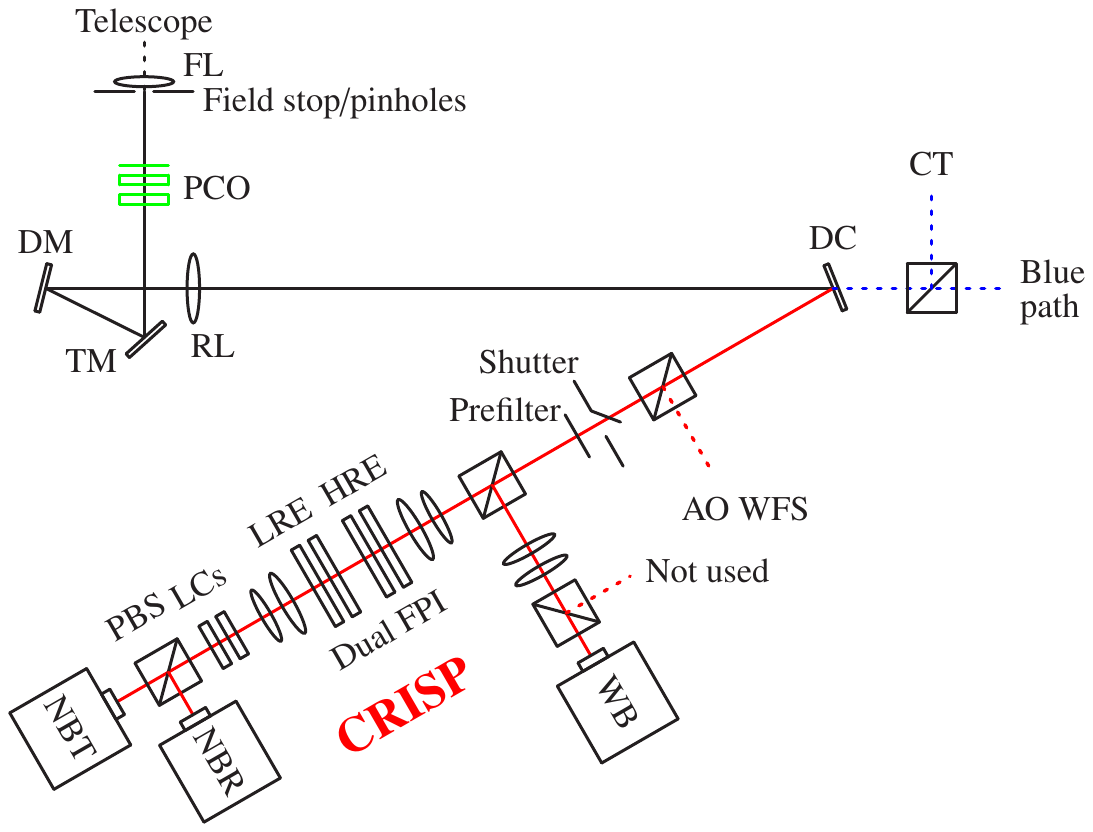}
  \caption{Sketch of optics, from bottom of telescope (inset A in
    Fig.~\ref{fig:sst}) to the CRISP. FL = Field Lens, PCO =
    Polarization Calibration Optics (not in the beam during normal
    observations), TM = Tip-tilt Mirror, DM = Deformable Mirror, RL =
    Reimaging Lens, DC = Dichroic Beamsplitter, CT = Correlation
    Tracker, AO WFS = Adaptive Optics WaveFront Sensor, FPI =
    Fabry--P\'erot Interferometer, HRE = High Resolution Etalons, LRE
    = Low Resolution Etalons, WB = Wideband, NBT = Narrowband
    Transmitted, NBR = Narrowband Reflected, LCs = Liquid Crystal
    modulators, PBS = Polarizing Beam Splitter.}
  \label{fig:opticalpath}
\end{figure}

We refer now to Fig.~\ref{fig:opticalpath}, where we show a schematic
view of the optical setup below the telescope, the details of which
are described in this section.

The field lens (FL) is near the focus from the Schupmann corrector and
it is designed to form a 34~mm diameter pupil image at the Adaptive
Optics (AO) deformable mirror (DM). Near the Schupmann focus, various
artificial targets and field stops can be slid into the beam. Just
outside the science FOV, a part of the beam is reflected toward a
wide-field wavefront sensor (WFS -- not shown in
Fig.~\ref{fig:opticalpath}), that can be used for monitoring the
seeing profile above the telescope \citep{scharmer10s-dimm+}.

The AO has two parts. The correlation tracker (CT) measures image
motion and uses the tip-tilt mirror (TM) to keep the image steady. The
recently installed 85-microlens Shack--Hartmann WFS measures the phase
of the wavefront at the pupil and tries to keep it flat by controlling
the shape of the 85-electrode monomorph DM from CILAS at a rate of
2~kHz. This system (Scharmer et al., in preparation) is similar to the
previous system by \cite{scharmer03adaptive}, although it is an
upgrade in both hardware and software.

Between the target slider and the TM, polarization calibration optics
(PCO) can be inserted in the beam. There are three parts: a red
filter, a rotatable linear polarizer, and a retarder. They are not
in the beam during normal science observations.

The TM and DM together make the beam horizontal and send it through
the reimaging lens (RL) toward the dichroic beamsplitter (DC), which
splits the beam (at 500 nm) into a red beam and a blue beam. The CRISP
is in the red beam, behind a beamsplitter that sends a small fraction
of the light to the AO WFS.

\subsection{CRISP}
\label{sec:crisp}
The tunable filter of the CRisp Imaging Spectropolarimeter
\citep[CRISP,][]{scharmer06comments,scharmer08crisp} is a dual
Fabry--P\'erot interferometer consisting of two etalons mounted in
tandem in a telecentric configuration. An etalon consists of two
parallel reflecting plates, forming a cavity where the light undergoes
multiple reflections. An infinite train of transmission peaks is
produced by multiple interference of light inside the cavity. The FPI
produces quasi-monochromatic images by combining a high spectral
resolution etalon (HRE) that defines the observed wavelength and a
lower spectral resolution etalon (LRE) that suppresses the first few
orders of secondary HRE transmission peaks. A prefilter limits the
spectral range of the light that passes through the optical system,
attenuating all higher order transmission peaks of both etalons
\citep[see][and references
therein]{1998thesos,2006cavallini,scharmer06comments}.
\begin{figure}[]
  \includegraphics[width=1.0\hsize]{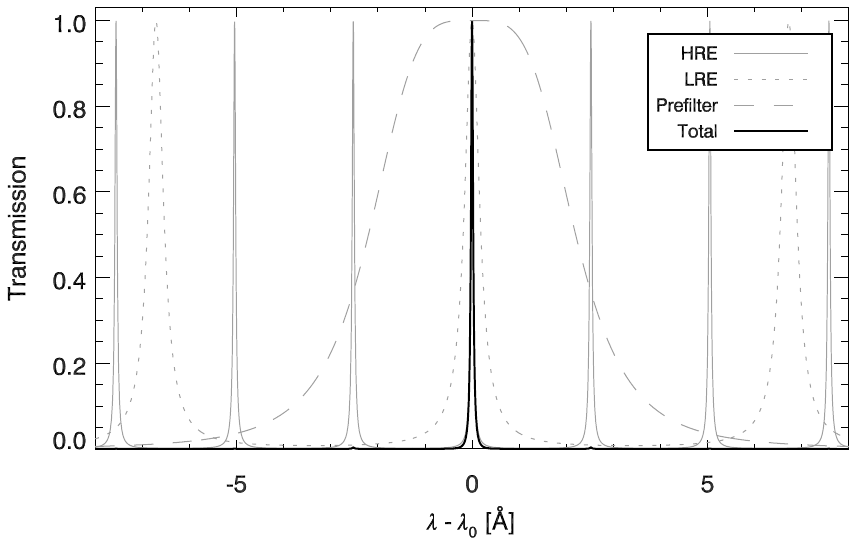}
  \caption{Approximate CRISP transmission profile 
    at 630~nm, computed theoretically assuming perpendicular incidence
    of the beam with the surface of both etalons. The individual 
    contributions from the high-resolution etalon, 
    the low-resolution etalon,
    and the prefilter 
    are shown along with the total transmission. All curves are
    normalized to unity.}
  \label{fig:crisptr}
\end{figure}

Figure \ref{fig:crisptr} shows the individual transmission curves for
the two etalons, the prefilter and the total CRISP transmission. Note
that the total transmission curve corresponds to that at the center of
the prefilter. Observations in the outer flanks of the prefilter
curve have larger contributions from the first few secondary
transmission peaks.

The transmission profile of each etalon can be approximately computed
for each ray as described by \citet{1998thesos},
\begin{equation}
  T(\delta\lambda, \delta R) 
  = \frac{1}{\displaystyle 1 + 
    \frac{4(R+\delta R)}{(1-R-\delta R)^2}
    \sin^2\Bigg(\frac{\Psi(\delta\lambda)}{2}\Bigg)},
  \label{eq:eta}
\end{equation}
where $R$ is the reflectivity of the etalon, and $\Psi$ is the phase
difference of succeeding reflections between the two reflecting plates:
\begin{equation}
  \Psi(\delta\lambda) = 2\cdot \bigg ( \frac{2\pi}{\lambda+\delta \lambda} \bigg) \cdot n \cdot D
  \cdot \cos\theta,\label{eq:phase}
\end{equation}
where $\lambda$ is the wavelength, $\delta \lambda$ is the cavity
error\footnote{The cavity error is really an error in the cavity
  separation $D$. Attributing it to the wavelength is mathematically
  equivalent ($\delta\lambda=\delta D\cdot \lambda/D$) and makes it
  immediately useful for correction of the data.}, $n$ is the
refractive index of the cavity, $D$ is the nominal separation between
the reflecting surfaces of the etalons, and $\theta$ is the angle of
incidence of the beam.

\begin{figure}[]
  \includegraphics[trim=0cm 4.7cm 0cm 0cm, clip=true, width=1.0\hsize]{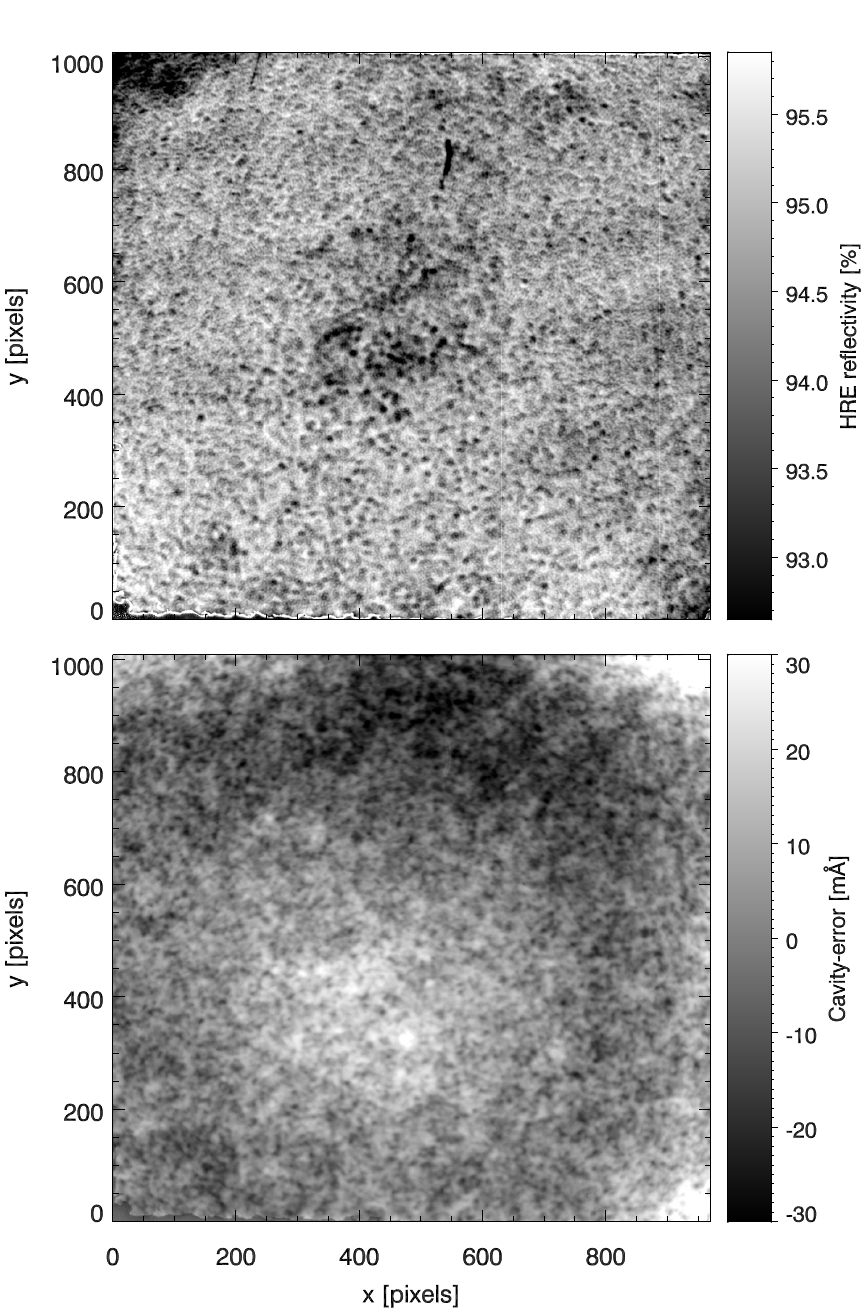}
  \caption{Characterization of the CRISP instrument at 617~nm.
    \emph{Top:} Reflectivity fluctuations of the high resolution
    etalon. \emph{Bottom:} Cavity error map of the high resolution etalon.}
  \label{fig:fit}
\end{figure}

The telecentric mount was chosen because it optimizes the image
quality in the focal plane \citep[see][]{scharmer06comments}. However,
this design introduces field-dependent variations in the transmission
profile:
\begin{itemize}
\item \emph{Imperfections on the surface of the etalons} influence the
  phase difference among successive reflections between the reflecting
  plates of the etalons. Thus, the position of the central wavelength
  of the transmission profile varies over the etalon surfaces,
  producing field-dependent wavelength shifts of the observed spectral
  lines. These are commonly known as \emph{cavity errors} (bottom
  panel in Fig.~\ref{fig:fit}).
\item \emph{Reflectivity variations} in the coating of the etalons
  make the width of the transmission profile to change along the FOV.
  Commonly known as \emph{reflectivity errors} (top panel in
  Fig.~\ref{fig:fit}).
\end{itemize}
We discuss in detail the characterization of CRISP parameters and the
treatment of cavity errors in Sect.~\ref{sec:char} and~\ref{sec:prob}.

The CRISP transmission profile depends critically on the reflectivity
of the coating of the etalon: narrower profiles correspond to higher
values of the reflectivity. Figure~\ref{fig:crispw} shows the full
width half maximum (FWHM) of the CRISP transmission profile as a
function of wavelength. To derive the FWHM, we have used the nominal
reflectivity values of the etalons as measured by the manufacturer and
Eq.~(\ref{eq:eta}).

Reimaging optics are placed before and after the etalons to make the
beam slower at the location of the etalons, and to place the focus
between the two etalons. After the second re-imaging, a pair of
modulators\footnote{The present modulators are a pair of nematic
  liquid crystals that change their states too slowly to finish during
  the covered phase of the chopper. They are therefore operated with
  overdrive, making them very sensitive to temperature variations
  within the observing room. New, faster, polychromatic modulators
  based on ferroelectric liquid crystals \citep{2011ASPC..437..413D}
  will be installed during the 2014 season.} produce four polarization
states that are linear combinations of the four Stokes parameters.
Finally, a linearly polarizing beam splitter separates the light into
horizontal and vertical polarized components and sends them to the
respective NB camera.

The CRISP instrument includes three Sarnoff CAM1M100 cameras with
back-illuminated, thinned CCD detectors consisting of 1024 by 1024
16$\times$16~\textmu{}m$^2$ pixels. One camera acquires wideband (WB)
images directly from the prefilter and two cameras acquire NB images
through the FPIs. After the prefilter wheel, a beam splitter sends
$\sim$8\% of the light to the WB channel and $\sim$92\% to the NB
channel. Placing the WB camera after the CRISP prefilter is crucial
for the MOMFBD image restoration that is described in
Sect.~\ref{sec:image-restoration}, because it relies on always having
simultaneous WB images with a wavelength near the NB wavelength.

The three cameras are synchronized by use of external shutter
(chopper), to facilitate image restoration. The chopper revolves at
$\sim$37~Hz with an exposure time of $\sim$17~ms, using the covered
$\sim$10~ms for read out.

\section{Newly developed methods}
\label{sec:new-developments}

In this section we describe methods of general use, that are original
to this publication or have not appeared in the peer-reviewed
literature.

\subsection{Flat-fielding and image restoration of wavelength
  inhomogeneous data}

\label{sec:prob}

In theory, corrections and calibrations to data should be applied
following the optical path in \emph{reversed order} (starting at the
detector). Accordingly, image restoration (see
Sect.~\ref{sec:image-restoration} below) should be the very last
correction to be applied to the data, after: dark current, flat-field,
polarimetric calibration, prefilter transmission and wavelength
shifts. The main problem is that we need to perform image restoration
before the polarimetric calibration to ensure that all narrow-band
states are perfectly co-aligned and de-stretched. We also need to
perform image restoration before compensating for wavelength shifts
(from the FPI) because otherwise atmospheric distortion would make the
spectra \emph{spatially incoherent} across the different line
positions, as solar features are moved around by the seeing. The data
processing described in this paper becomes highly complicated because
we need to perform image restoration at earlier stages than we should,
and cavity errors cannot be corrected prior to image-restoration.

Conceptually, it should be possible to perform a forward model fitting
of the spectral line in each position within the FOV, including
polarimetry as well as wavefront aberrations and geometrical
distortions from anisoplanatism, and try to reproduce the observed
data. In practice, this kind of modelling is very difficult to
implement and the computations are likely to be demanding. Perhaps the
techniques described by \citet{2012A&A...548A...5V} set the ground for
further developments in this direction.

Current implementations of image restoration deal with
quasi-monochromatic data, where the distortions of the object (the
Sun) are assumed to be due to wavefront aberrations only. The combined
effect of cavity-errors and a line profile imprints the images with a
pattern of intensity fluctuations, resulting from the wavelength shift
of the transmission profile, violating this assumption. The effect is
strong enough to disturb the wavefront sensing process, particularly
in bad seeing. Deconvolution of the cavity-error pattern causes
further artifacts in the restored images.

Without any kind of wavelength shift compensation, the restored images
show a background of strong artifacts that are somewhat correlated
with the distribution of wavelength shifts. This pattern is
particularly visible when the gradient of the spectral profile is
steep, and it disappears in the continuum. These are the issues that
make the data reduction process entangled and complicated for
instruments like CRISP and TESOS.

\citet{2011schnerr} proposed an approximate solution: once the
cavity-error map is known, one can use the quiet-Sun average profile,
present in the flat field data, to approximately correct the effects
of cavity-errors, assuming that the slope of the resolved
pixel-to-pixel profile is somewhat similar to that from the quiet-Sun
average. This approach allowed to easily compute a correction that
makes the image smoother for the image restoration process. Afterward,
the restored images would be multiplied by the ratio between this
incorrect flat-field and the cavity-error free one (where the
quiet-Sun profile is removed).
 
Obviously, there are limitations to this approach. A spectrum from a
sunspot is hardly similar to a quiet-Sun mean profile, in strength,
width and shape. But this approximation seems to work well when the
seeing conditions are excellent, probably because the contrast of the
images dominates over the intensity fluctuations introduced by cavity
errors. Normally, the effect of cavity errors is stronger at
\emph{shorter} wavelengths, because the ratio $D/\lambda$ in
Eq.~(\ref{eq:phase}) is \emph{larger}.
\begin{figure*}[]
  \centering
  \includegraphics[width=\hsize, trim=3mm 0 0 0, clip]{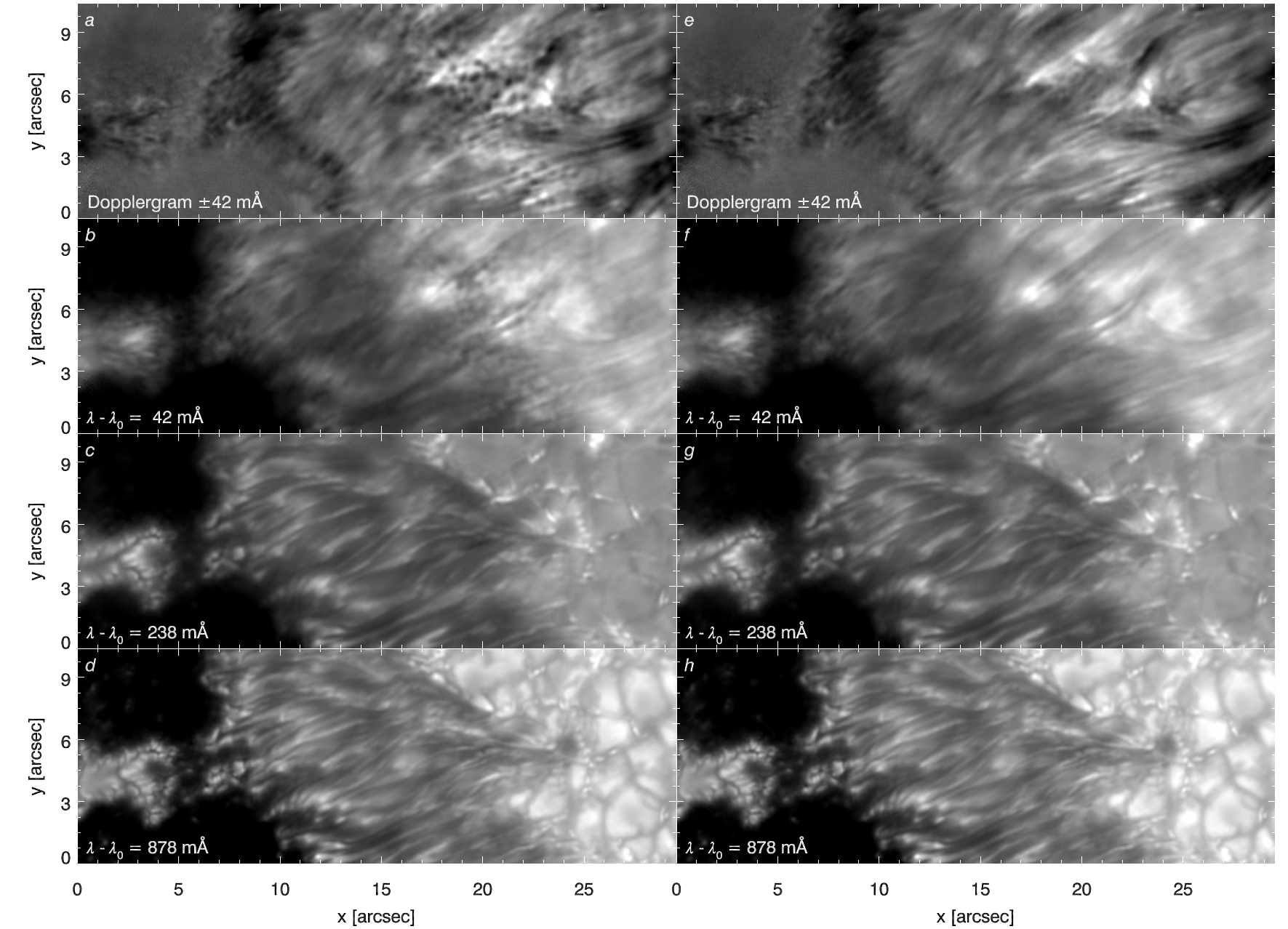}
  \caption{Observations in the \ion{Mg}{i}~$\lambda 5173$ line. The
    top row correponds to Dopplergrams (calculated from images at $\pm
    42$~m\AA). The other panels, in descending order, show red wing
    images, $\lambda-\lambda_0=42$, 238, and~878~m\AA, respectively.
    The images in the left column were restored with the flat-fielding
    scheme by \citet{2011schnerr}. Artifacts due to cavity errors
    appear as small-scale grains, clearly visible in panels $a$ and
    $b$. In contrast, no such errors are visible in the right column,
    corresponding to images restored with our new method.}
  \label{fig:cerrmg}
\end{figure*}

In Fig.~\ref{fig:cerrmg} we illustrate the effect with a CRISP dataset
acquired in the \ion{Mg}{i}~$5173$ line, where the flat-fielding
scheme proposed by \citet{2011schnerr} leads to small-scale artifacts
in some restored images (left column), especially close to line center
where the line profile is very steep (panel $b$ and the Dopplergram in
panel $a$). Note that artifacts will preferentially appear in the
presence of a steep line profile and low image contrast and therefore
not necessarily in all line positions (e.g., panels $c$ and $d$) or
even within the entire line scan.

\begin{figure}[]
  \centering
  \includegraphics[width=\hsize]{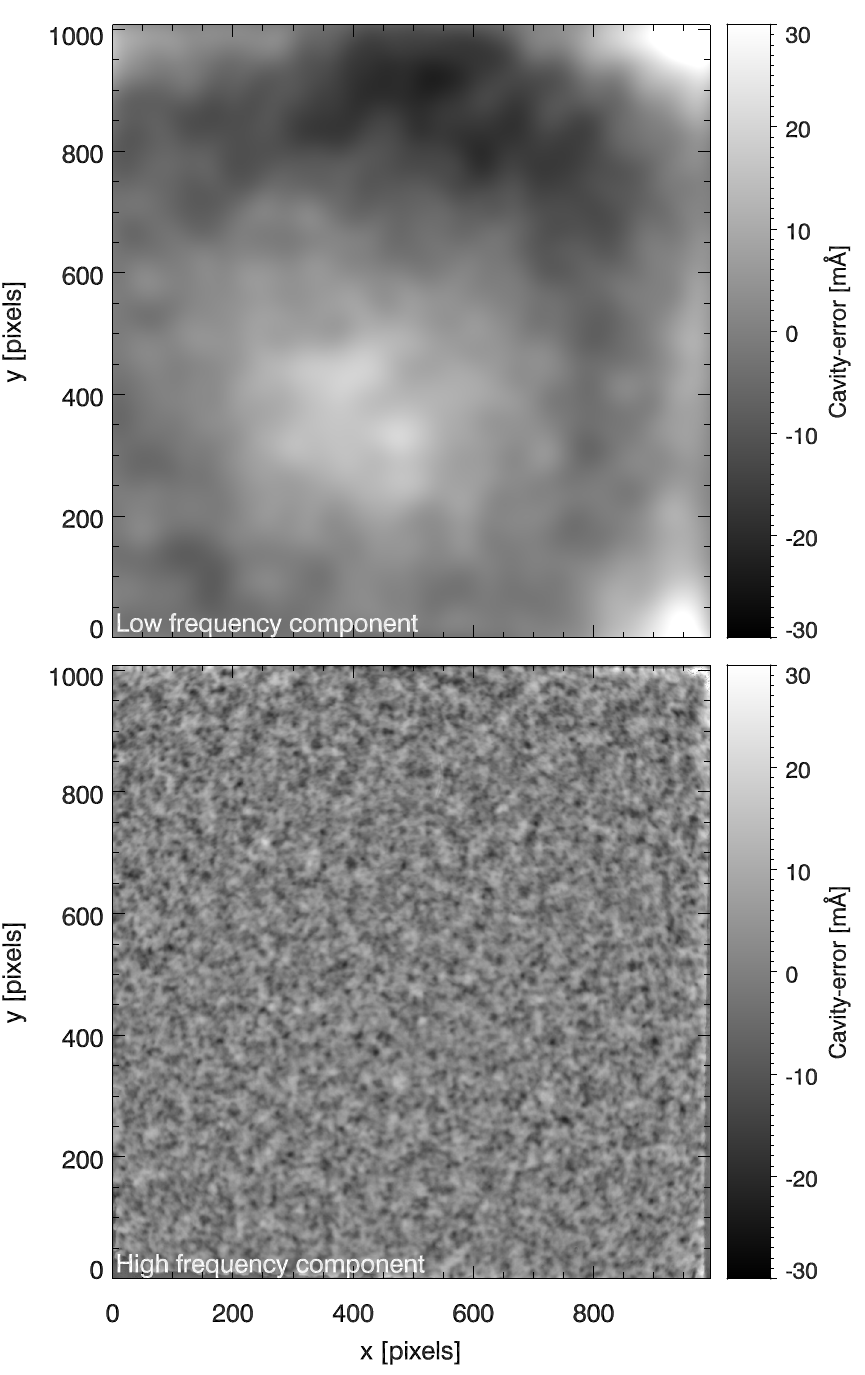}
  \caption{The cavity error map from Fig.~\ref{fig:fit}, decomposed
    into low spatial frequencies (top) and high spatial frequencies
    (bottom).}
  \label{fig:cmap}
\end{figure}

In this study, we present a more accurate way to deal with the
flat-fielding of non-monochromatic images for image restoration. The
idea is to use non-restored data to measure the observed spectra
across the FOV, and then calculate the corrections needed to
compensate for intensity fluctuations introduced by the cavity error.
For each line scan, we add all the images acquired in a burst for each
wavelength, resulting in a data-cube with line profiles for each
pixel, $I(\lambda,x,y)$.

These profiles can be shifted by interpolation, producing the
data-cube $I^*\bigl(\lambda-\delta\lambda(x,y),x,y\bigr)$, where
$\delta\lambda(x,y)$ is a cavity map, estimated as in
Sect.~\ref{sec:char} below (see also Fig.~\ref{fig:fit}). The
corrected flat-field, that will be used to perform the image
processing, $F_\text{corr}$, can then be constructed using the
cavity-error free flat, $F_\text{cef}$, multiplied by the ratio
between the shifted and unshifted profiles,
\begin{equation}
  F_\text{corr} (\lambda,x,y)
  = 
  F_\text{cef}(\lambda,x,y) 
  \frac{I(\lambda,x,y)}{I^*\bigl(\lambda-\delta\lambda(x,y),x,y\bigr)}.
  \label{eq:cerr}
\end{equation}
This method should provide cavity-error free images because it
effectively shifts the observed line profile so that the wavelength is
constant within a frame. In our tests, however, we noticed that some
residual distortion could be introduced in the restored images because
the spectra from our summed (unprocessed) data are not free of
atmospheric distortions.

To overcome this inconvenience, a mathematical trick can be used
(Scharmer 2014, private communication). Due to the limited spatial
extent (a few arcsec) of the PSF, the image restoration is only
affected by the presence of fluctuations at high spatial frequencies,
whereas it is rather insensitive to the presence of large scale
variations. This strongly limits the distance over which intensity
information from different wavelengths are mixed. Thus, it is the
small-scale fluctuations that cause artifacts in the deconvolution
because they are \emph{discontinuous} to a certain extent, whereas the
large-scale fluctuations do not introduce this problem.

By correcting only the high frequency fluctuations, the profile is
shifted by a significantly smaller amount and potential residual
distortions remain invisible. The low frequency component can be
corrected, once the images have been restored from wavefront
aberrations. Also, the correction proposed in Eq.~(\ref{eq:cerr}) is
strictly more appropriate since we are assuming that the slope of the
line profile at any given wavelength has linear dependence. This
assumption becomes less accurate when corrections are large.

We illustrate the concept in Fig.~\ref{fig:cmap}, which shows the
cavity error map from Fig.~\ref{fig:fit}, decomposed into two
components: the low spatial frequencies (9.16~m\AA\ amplitude RMS
contrast) and the high spatial frequencies (5.07~m\AA\ amplitide RMS
contrast). Both panels are scaled between $\Delta\lambda = \pm
30$~m\AA. Note the much smaller amplitude of the high frequency
component. We show the results of a restoration performed with our new
method in the right column of Fig.~\ref{fig:cerrmg}. The Dopplergram
and all images are free from visible cavity-error artifacts.

\subsection{Pinhole calibration}
\label{sec:pinhole-calibration}

\begin{figure}[tb] 
  \centering
  \includegraphics[width=0.9\linewidth]{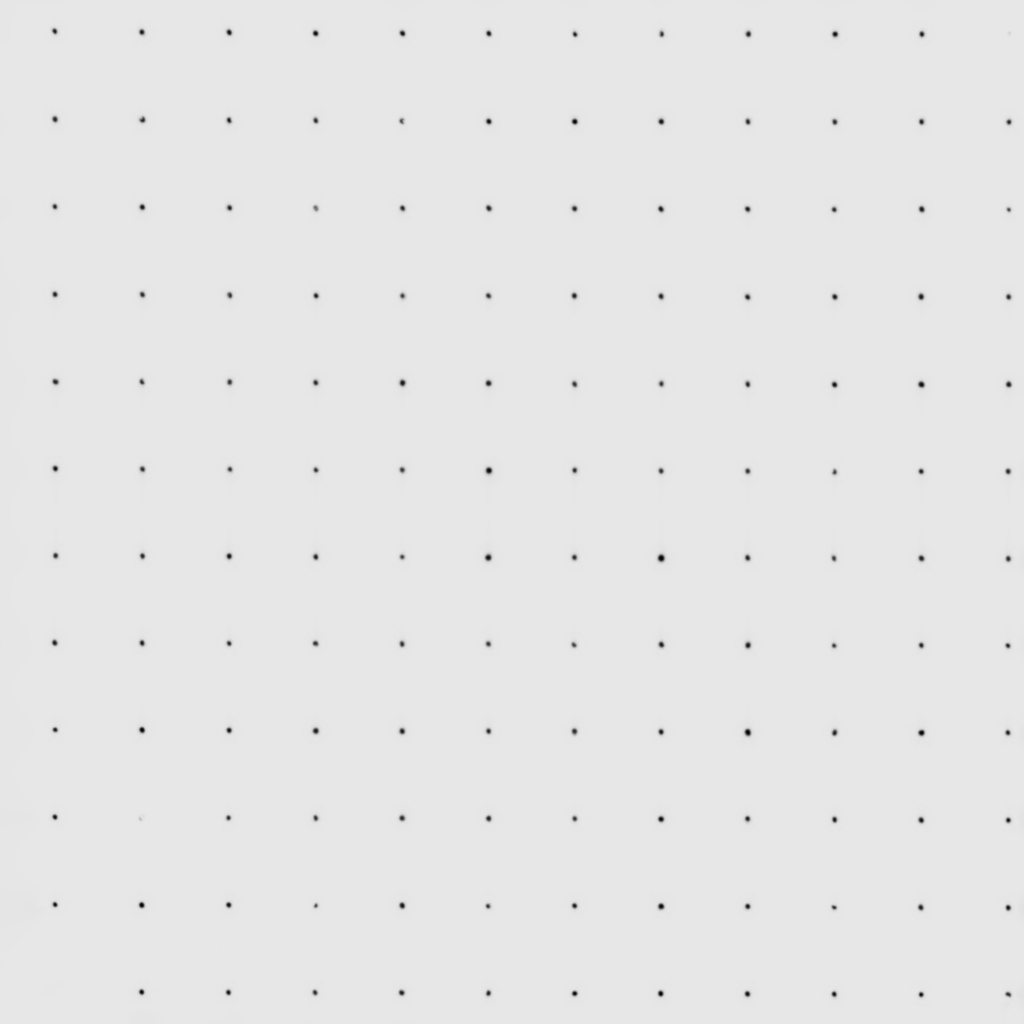}
  \caption{A 1024$\times$1024-pixel
    (60\protect\farcs4$\times$60\protect\farcs4) image of the pinhole
    array, inverse intensity scale.}
  \label{fig:pinhole-array}
\end{figure}

As reference for inter-camera alignment, we use a pinhole array that
is mounted at the Schupmann focus. Most telescopes should have an
early focus where such a target can be mounted. {Multiple images} of this array
are routinely
collected every observing day {in the same way as flat field
  data, and co-added separately for each camera and state}. A grid of about 12$\times$12 pinholes
fit in the detector FOV, see Fig.~\ref{fig:pinhole-array}. Removing
the pinholes closest to the edges leaves about 10$\times$10 pinholes
for further processing.

The maximum intensities of the pinholes are first used for finding out
the relative orientation (rotation and mirroring) of the cameras, as
well as alignment errors on the order of the grid spacing or larger
(which should never happen). This works because the amount of light
passing through the different holes depends on size and blockage by
dust or dirt, forming a distinct pattern.

{The summed pinhole images have negligible solar structure within the
small area of the holes, and therefore} the object {can be
assumed to be} the same in all
cameras and states. By individually MOMFBD-processing subfields
centered on each of the pinholes, we estimate field-dependent
alignment parameters that represent the alignment as a function of
position in the FOV. Two-dimensional fits to these estimated values
are stored to disk, where the MOMFBD program can look up the shifts
needed to align the cameras at a particular location.

Over the last 10 years, the non-MOMFBD parts of this procedure have
been implemented in ANA, shell script, python, and C++. When we
re-wrote it in IDL, we also upgraded the procedure slightly.

Some pinholes are completely or partially blocked, giving bad MOMFBD
inversions, see gaps in the pinhole grid in
Fig.~\ref{fig:pinhole-array}. Therefore we process only pinholes
brighter than a threshold, reducing the number of outliers included in
the fit.

Two problems with the pinhole processing were solved by methods
resembling what was done by \citet{lofdahl12sources}, who were able to
measure not only focus but also higher order aberrations from pinhole
data. We now sum the pinholes with sub-pixel alignment after filling
in bad pixels with interpolation. This solves the problem with
sub-pixel jitter between exposures, which can violate the assumption
of a common object if the sums do not involve the same exposures for
all cameras. We then refine the dark level of each subfield
individually. When all the energy in an image is concentrated to a few
pixels, also small errors in the dark level can make problems for the
assumption of a common object. Using the WB as the \emph{anchor}
channel, in the old procedure, pinhole MOMFBD was run separately for
WB+NBT and WB+NBR, respectively. Now we run WB+NBT+NBT, resulting in
(at least potentially) more constrained inversions.

These changes improved the consistency of the estimated offset
surfaces between different states. This strengthened the case for
another change in the procedure. Earlier, we have mostly measured the
NB pinhole shifts with respect to the WB, not only for each NB camera,
but for each combination of camera, prefilter, wavelength tuning, and
modulator state. This requires very large amounts of pinhole data to
be collected for observing sequences with many wavelength points. The
pipeline will still perform the pinhole calibration for all data that
are collected but the recommendation is now to collect data only for a
single state in each scan (prefilter), and then use the measured
offsets for all states. An exception will be made for the very
broadest lines, where data should be collected for a few wavelength
tunings and the offsets then interpolated.

\subsection{Image scale}
\label{sec:image-scale}

Changes on the optical table can change the image scale significantly.
We also discovered recently that the CRISP prefilters have a slight
optical power, enough to move the focus by a few mm (optionally
compensated for when observing with an extra focus term on the DM) and
also change the image scale slightly. The pinhole array mounted at the
Shupmann focus, used as a reference for camera alignment (see
Fig.~\ref{fig:pinhole-array}), is therefore used as the reference also
for determining the image scale in angular units per pixel.

\begin{figure}[tb]
  \centering
  \includegraphics[width=.495\linewidth]{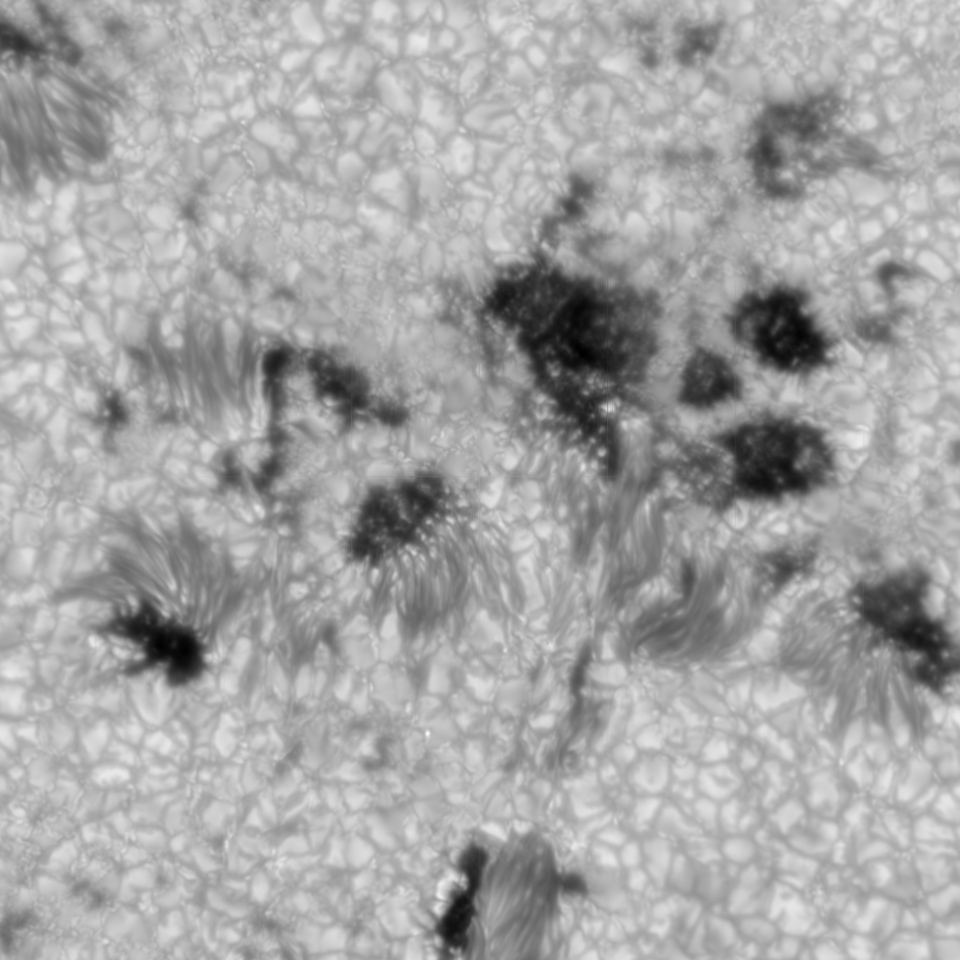}
  \hfill
  \includegraphics[width=.495\linewidth]{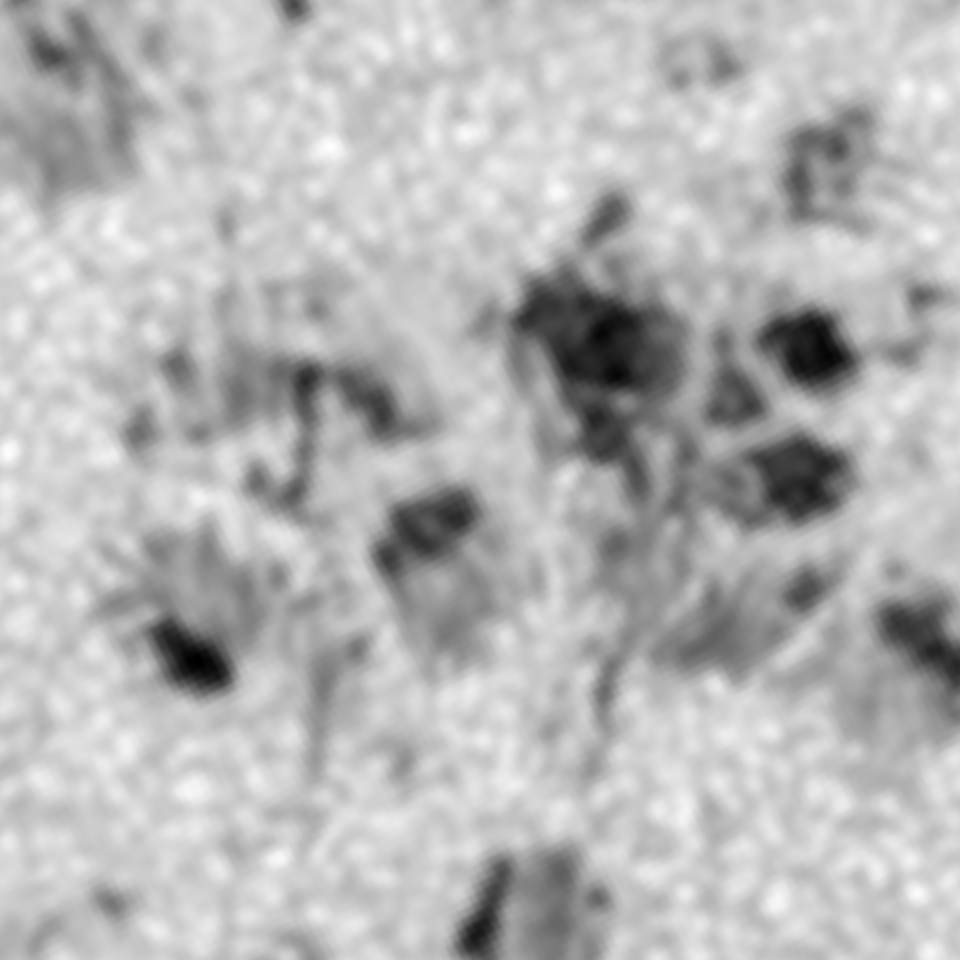}
  \caption{Co-temporal and co-spatial 558~nm SST/CRISP (left) and
    617~nm SDO/HMI (right) images used for image scale measurements.
    The FOV covers part of AR1589, which was near disk center on
    2012-10-15.}
  \label{fig:sst-hmi-images}
\end{figure}

To serve as such a reference, the pinhole grid spacing must be known
in angular units. By use of a code developed by
\citet{noren13calibration} we measured the image scale in the CRISP WB
camera by comparison with approximately co-temporal 617~nm continuum
images from SDO/HMI \citep{2012SoPh..275..229S}. Images of AR1589, an
active region with many high-contrast features distributed over the
SST FOV, were collected on 2012-10-15 and MFBD-restored in groups of
$\sim$100 exposure, then down-sampled to the same resolution as the
HMI images (sample images are shown in Fig.~\ref{fig:sst-hmi-images}).
We measured in two wavelengths, through the 558 and 630 nm CRISP
prefilters, and got image scales 0\farcs05900/pixel and
0\farcs05876/pixel, respectively. The accuracy of these measurements
is determined by two factors. The HMI image scale is known to a
relative error of 0.07\% or better, mostly limited by the value
adopted for the solar radius \citep{schou13}. The standard deviations
of the image scale measurements performed separately on about 150 SST
images is on the same order of magnitude, resulting in a combined
uncertainty of about 0.1\%.

As part of the CRISPRED pipeline, we measure the pinhole spacing (in
pixels) by measuring the individual positions of each pinhole in the
FOV and fitting (using MPFIT) those positions to a model with five
parameters: the coordinates of the pinhole nearest the origin, the
grid spacing in the two grid axis directions, and a rotation angle.
Assuming we can measure the pinhole positions to better than pixel
accuracy, over the $\sim$1000 pixel detector size we should get
accuracy that is significantly better than the 0.1\% of the SST/HMI
image scale calibration.

Measuring the pinhole grid spacing in pixels in pinhole array images,
collected with the same cameras and on the same day as the active
region data above, gives 86.75 pixels and 87.04 pixels, respectively,
in the two wavelengths. This means that the measurements agree on a
grid spacing of 5\farcs12 ($5\farcs116\pm0\farcs005$) for both
wavelengths. This number is used in the pipeline as the reference for
calculating the image scale in all cameras. The so calibrated image
scale is then used by the MOMFBD program (Sect.~\ref{sec:momfbd}) and
should also be used for interpretation of the data.

\section{The CRISPRED pipeline}
\label{sec:data-reduction-steps}

In this section we describe the corrections involved in the data
reduction of CRISP, but most of them are common to instruments with
similar characteristics.

\begin{figure*}[p]
  \centering
  \includegraphics[width=\hsize]{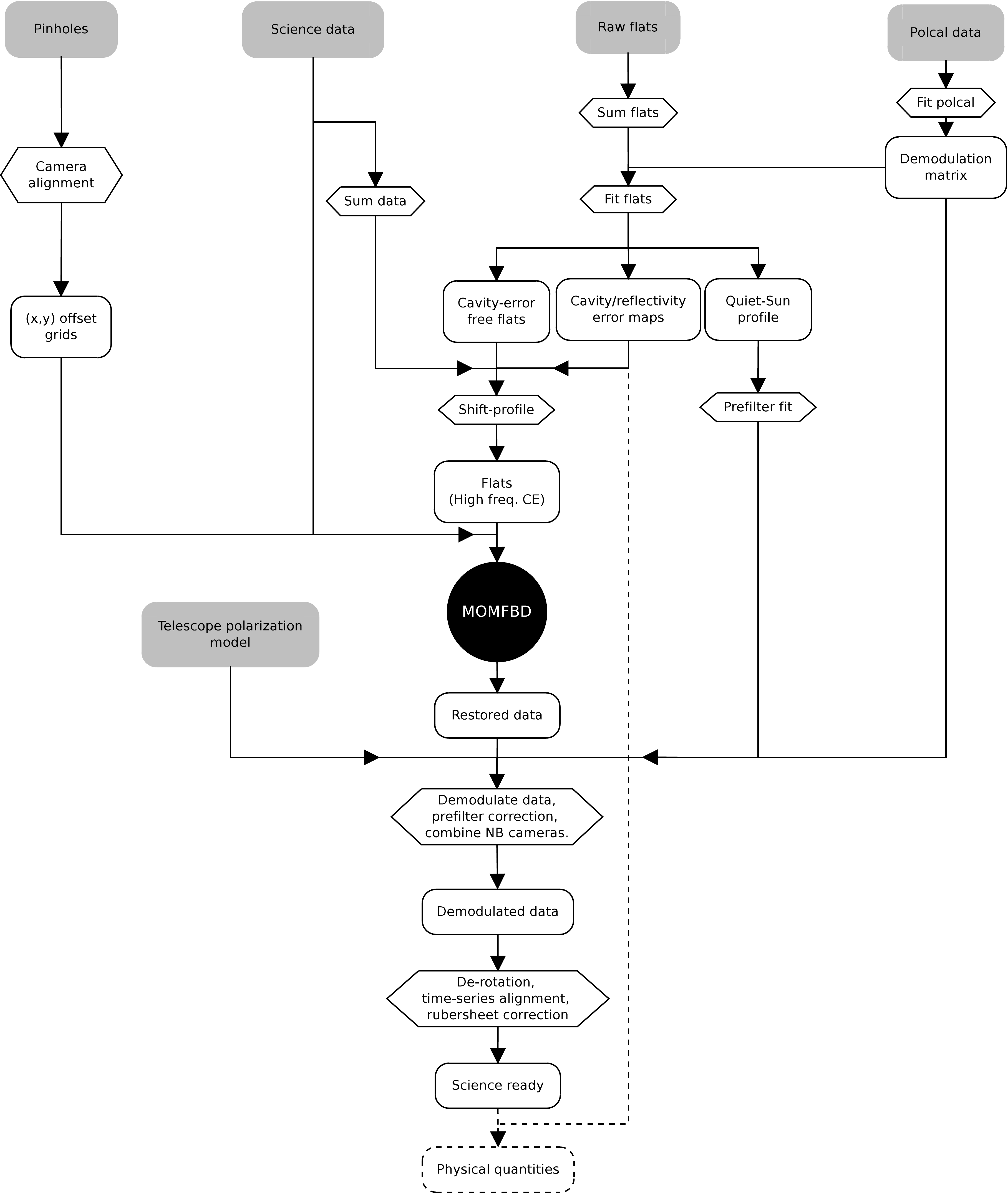}
  \caption{Data flow in the CRISPRED pipeline. Starting points are the
    rounded boxes with a gray background: the raw data and the
    telescope polarization model. The diamond boxes indicate fitting
    and other processing. Products from fitting routines or data
    processing are represented by rounded, white boxes. The distinct
    symbol for the MOMFBD processing indicates that it is not done in
    IDL. Reflectivity and cavity errors must be included in the
    interpretation of the science data (outside the CRISPRED pipeline,
    hence indicated with dashed lines). Omitted in the figure for
    brevity, all science and calibration images are corrected for dark
    current.}
  \label{fig:flow}
\end{figure*}

Figure~\ref{fig:flow} illustrates the data flow of the CRISPRED
pipeline. All the details are provided in the following sections, but
we include here a very brief summary of the data flow.

The following effects are compensated for within the pipeline. Some
effects are coupled as argued in Sect.~\ref{sec:prob}, which makes the
processing more complicated.
\begin{enumerate}
\item Spatial variation in detector gain and zero level offset (bias),
  IR detector transparency.
\item Spatial variation in the CRISP FPI passband profiles;
  reflectivity, cavity errors.
\item Optical aberrations from atmospheric turbulence (seeing) as well
  as from the optics.
\item Inter-camera misalignments.
\item Field-dependent and time-varying instrumental polarization.
\item Image field rotation, temporal misalignment and rubber-sheet
  warping from seeing.
\end{enumerate}

The first steps in a typical reduction are to perform the polarimetric
calibration~(Sect.\@~\ref{sec:polcal}) to derive a demodulation matrix
for each pixel. The latter are used to demodulate the flat-field data.
We prepare our flat-field images (Sect.~\ref{sec:detector-effects})
using spectra from real observations and our processed flats
(Sect.\@~\ref{sec:prob}). We use the Stokes~$I$ flat-field images to
characterize CRISP (Sect.\@~\ref{sec:char}).

Our two NB cameras are aligned to the WB camera by use of pinhole
images (Sect.\@~\ref{sec:pinhole-calibration}), that allow the
estimation of offsets in $(x,y)$ positions. These offsets are used by
the MOMFBD program in order to make restored images that are well
aligned (Sect.\@~\ref{sec:momfbd}).

The restored images are demodulated and compensated for
telescope-induced polarization (see Sect.~\ref{sec:polcal}). Later, we
estimate and apply corrections to obtain a homogeneous time-series
(Sect.\@~\ref{sec:time}).

\subsection{Polarimetric calibration}
\label{sec:polcal}

Above the position of the PCO, the SST has three lenses and four
mirrors that modify the polarization properties of the light as a
function of {pointing angle (and introduce image rotation, see
  \S\ref{sec:time})}. Additionally, the optical table where the CRISP
instrument is mounted contains a train of optical elements that also
affect polarization (see Fig.~\ref{fig:opticalpath}). We use two
modulators to produce four polarization states that are linear
combinations of the wanted Stokes parameters. A linearly polarizing
beam splitter is used to split the beam in two orthogonal states that
are detected simultaneously by each of the NB cameras. This dual-beam
setup is necessary for removing, to first order, seeing-induced
polarization cross-talk \citep[see][and references
therein]{2012ApJ...757...45C}. For each camera and wavelength, and in
each $(x,y)$ pixel, the intensity for the four polarization states can
be written as
\begin{equation}
  \bm{I}_\text{obs}(x,y) =
  \textbf{M}_\text{table}(x,y) \, \textbf{M}_\text{tel}(\theta, \varphi) \, \bm{I}_\text{real}(x,y),
\end{equation}
where $\bm{I}_\text{obs}$ is the detected intensity vector,
$\textbf{M}_\text{tel}$ is the telescope matrix (which depends on the
azimuth $\varphi$ and the elevation $\theta$ of the telescope
pointing) and $\textbf{M}_\text{table}$ is the modulation matrix and
$\bm{I}_\text{real}$ is the real Stokes vector ($I,Q,U,V$).
\citet{selbing05sst} proposed a telescope polarization model for the
SST and provided a telescope matrix $\textbf{M}_\text{tel}$ for
630~nm. We use a similar model and up-to-date parameters have been
acquired for the prefilters relevant for polarimetry as indicated in
Table \ref{tab:pref}.

The SST polarization model is constructed using
the Mueller matrices of each polarizing element,
\begin{equation}
  \begin{split}
    \textbf{M}_\text{tel}\left(\theta, \varphi \right)  
    & = \textbf{R}_{f+}(a_{11}) \cdot \textbf{N}_{s}(a_9,a_{10})\cdot \textbf{N}_{f}(a_7,a_8) \\
    & \quad \cdot \textbf{R}_{f-}(a_{11})\cdot \textbf{R}_\text{az}(\varphi)
    \cdot \textbf{N}_\text{az}(a_5,a_6) \cdot \textbf{R}_\text{el}(\theta)\\ 
    & \quad \cdot \textbf{N}_\text{el}(a_5,a_6) \cdot \textbf{L}(a_0,a_1,a_2,a_3,a_4). \label{eq:telmod}
  \end{split}
\end{equation}

In this model, the entrance lens $(\textbf{L})$ is a composite of
retarders that modulate the light (5 free parameters). Mirrors are
noted with $\textbf{N}$ and have two free parameters. Rotations to a
new coordinate frame are indicated with $\textbf{R}$. The exact form
of these matrices is given by \citet{selbing05sst}. The properties of
the optical elements are wavelength dependent and are contained in the
model parameters $a_0,a_1,\ldots,a_{11}$.

The polarimetric calibration of the optical table is performed as
described by \citet{2008vannoort}. To acquire the calibration data,
known polarization states are generated using a linear polarizer and a
quarter-wave plate (QWP hereafter), located just after the exit window
of the telescope (PCO in Fig.~\ref{fig:opticalpath}). Typically, for
each linear-polarizer angle, the QWP rotates $360\degr$ in $10\degr$
steps. Two linear-polarizer angles suffice to accurately infer the
modulation matrix $\textbf{M}_\text{table}$. The presence of
polarizing optics close to some of the focal planes introduces
significant field-dependent fluctuations in the elements of the
matrix. To account for them, the elements of $\textbf{M}_\text{table}$
are estimated (fitted) for each pixel on the CCD using a
multi-threaded C module.

After the MOMFBD process, blurred versions of the demodulation matrix
(the inverse of the modulation matrix $\textbf{M}_\text{table}$) {are}
applied to the restored images on a pixel by pixel basis as described
by \citet{2011schnerr}. Basically, for each NB camera, the
demodulation matrix is clipped in the same subfields as the images,
then convolved with the average seeing PSF of the corresponding patch
and, finally, stitched together in the same way as the restored
images. Before applying the demodulation matrix to the data, residual
seeing distortions are removed from each of the polarization states as
described by \citet{henriques12three-dimensional}.

Data from the two NB cameras are combined after demodulation and
normalization with respect to the mean intensity (Stokes I), so each
contribute with the same weight to the average. This step is necessary
because the cameras do not have strictly the same gain.

Telescope polarization is compensated for in a subsequent step, if the
parameters of the telescope model are known for the observed spectral
region.

\subsection{Detector effects}
\label{sec:detector-effects}

The pixel by pixel gain and bias are in principle easy to calibrate.
Exposures collected while the beam is blocked (dark frames) establish
the bias and flat fields establish the gain. The off-disk sky being
too dark for solar setups, flats in solar telescopes are collected as
averages of quiet-sun granulation. For the SST, this is done while
moving the telescope pointing in a circle over quiet Sun near disk
center and blurring the image by putting out random voltages on the
DM. Flats are collected individually for each wavelength tuning and
polarization state.

Both darks and flats are made by averaging many exposures in order to
reduce shot noise and -- in the case of flats -- reduce the influence
of solar structures. We check statistics for all dark and flat frames
and remove outliers before summing. This way we essentially remove the
risk of having the final flats and darks being destroyed by the
occasional camera or shutter glitch, birds flying in front of the
telescope aperture, etc.

Bad pixels are zeroed in the gain tables to be read by the MOMFBD
program, which interprets the zero gain pixels as instructions to
calculate interpolated values for the corresponding pixels in the
image data. Isolated or small groups of bad pixels are found with
heuristics based on identifying outliers in a unsharp-masked version
of the summed flat fields.

Large CCDs are often divided into regions that are read out
separately. We have found that the borders of such regions can cause
artifacts in the image restoration, probably due to slight
non-linearities that differ between the regions. This is the case for
the Sarnoff cameras in the CRISP setup. We routinely mark a column of
pixels as bad at such borders, which appears to be an easy remedy for
such problems. Because of this, the absolute majority of pixels to be
interpolated by the MOMFBD program are oriented in columns. We have
therefore implemented an option to the MOMFBD program, to use
1-dimensional interpolation across those columns instead of the
default circularly symmetrical interpolation. We have found that this
further reduces the artifacts from such borders.

A WB observed image is corrected by pixel-by-pixel subtraction of the
bias and division with the gain. However, the flats correct not only
for detector characteristics but also for imperfections in the
near-focus optical elements, e.g. dust. These effects couple with
polarization and are therefore time-dependent for the NB images. The
procedure for dealing with this is described in Sect.~\ref{sec:prob}
above.

At near-IR wavelengths, the silicon in the detector substrate can
become semi-transparent, so that back-scatter from electronics and
other camera parts behind the detector adds a component to images that
cannot be removed by normal flat-fielding. We have adopted the
procedure for removing such effects described by
\citet{delacruz10measuring} and \citet[Appendix A.1]{2013delacruz}.
The back scatter is modeled with a PSF describing the scattering in
the material behind the CCD and a gain factor representing the amount
of light that returns from that direction. For our Sarnoff cameras,
this procedure is needed at wavelengths longer than $\sim$700~nm
(currently affecting data acquired through the 777~nm and 854~nm CRISP
prefilters).

\subsection{FPI characterization}
\label{sec:char}

In this section we summarize the methods used to characterize the
cavity-error and reflectivity-error maps, and how to estimate and
remove the quiet-Sun profile from the flats.

The way flat-fields (flats) are made, by averaging many short-exposure
images collected while the telescope moves in small circles around the
center of the solar disk, allows the assumption that the spectrum
present at each pixel corresponds to a spatially averaged quiet-Sun
profile at disk center. Therefore, the flats are also affected by the
presence of cavity errors, although with a different line profile than
line profiles in the observations.

We use a self-consistent method developed by \citet{2011schnerr}. The
basic idea is to remove the imprint of spectral lines from the summed
flats. The presence of cavity and reflectivity errors, fringes and
variations of the prefilter across the FOV, can be understood as a
distortion of the profiles, and it can be modeled using a
self-consistent iterative scheme, here implemented in an external
multi-threaded C++ module that makes use of a Levenberg--Marquardt
algorithm to fit the data \citep[cMPFIT,][]{2009craig}.

In the first iteration, cavity errors are unknown, so we estimate the
quiet-Sun profile by fitting a cubic non-overshooting Bezier-spline
\citep[see summary by][]{2003auer} to the average spectrum imprinted
in the flats. That average spectrum is shifted in wavelength and
scaled in intensity to match the observed spectrum at each pixel,
providing a first estimate of the cavity-error map and the gain at
each pixel. For stability purposes, in this first iteration, the
prefilter variations and the reflectivity are assumed to be zero.

Using the fitted values, we place all spectra (across the FOV) in a
large array of $(\lambda,I)$ values, corrected for the cavity error
shift and gain value. A new estimate of the quiet-Sun average is
derived by fitting a Bezier-spline to all data points. In the
following iterations, a slightly more complete model is used to fit
the data, including the gain value, a cavity-error, reflectivity
fluctuations and prefilter/fringe variations. The latter are fitted
using a polynomial multiplicative term to the average spectrum. The
model is summarized as follows:
\begin{equation}
  I(\lambda) = 
  G \cdot \mathfrak{F}^{-1} \bigg \{ \mathfrak{F}\{ \hat{I}(\lambda) \}
  \frac{\mathfrak{F}\{ T(\delta\lambda,\delta R)\}}{\mathfrak{F}\{T(0,0) \}} \bigg \} 
  \bigg (1+\sum_{n=1}^N c_n\lambda^n \bigg ),
\end{equation}
where $G$ is the value of the gain, $\hat{I}$ is the estimate of the
quiet-Sun average, $T(\delta\lambda,\delta R)$ is the combined FPI
transmission profiles of the two etalons (see Eq.~(\ref{eq:eta})),
$\delta \lambda$ and $\delta R$ are the cavity and reflectivity errors
respectively, and $c_n$ are the coefficients of the multiplicative
polynomial component. All the fitted parameters are allowed to vary
between pixels. Note that $T(0,0)$ is a transmission profile computed
using the reflectivity value provided by the FPI manufacturer and no
shift. Also note that the length of the polynomial component can be
adapted to each case, depending on the wavelength coverage of the
observations and how strong fringes are at that wavelength. The
$\mathfrak{F}$ operator represents the one-dimensional Fourier
transform and $\mathfrak{F}^{-1}$ its inverse. At each pixel, we fit
the value of $\delta \lambda$, $\delta R$ and $c_1$,...,$c_n$. Further
details are given by \citet{2011schnerr}.

\begin{figure}[]
  \includegraphics[trim=0cm 0cm 0cm 13.4cm, clip=true, width=1.0\hsize]{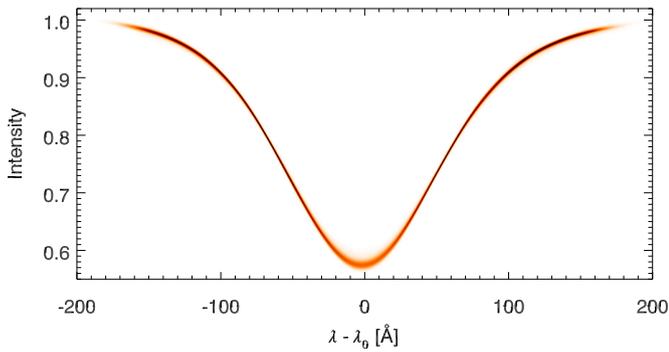}
  \caption{Two-dimensional histogram of the quiet-Sun spectra at the
    \ion{Fe}{i}~$617.3$~nm line, from the flat
    field data, where each spectrum (from a different pixel) has been
    shifted to correct the cavity error and scaled according to the
    gain value.}
  \label{fig:fit2}
\end{figure}

Figure~\ref{fig:fit} illustrates the resulting reflectivity (top) and
cavity (middle) error maps that are inferred using our routines.
Additionally, in Fig.~\ref{fig:fit2} we represent a two-dimensional
histogram computed from all spectra in the FOV, corrected for cavity
errors, normalized by the gain value and corrected for the polynomial
component. In this figure we have not corrected the data for
reflectivity errors, although those are included in our model. Once
all the corrections are applied, we expect all data points to tightly
describe the same spectral profile.

Chromospheric lines, like \ion{H}{i}~$\lambda6563$ and
\ion{Ca}{ii}~$\lambda8542$, are much wider than photospheric lines.
Under those circumstances, if the slope of the line has a linear
behavior within the width of the transmission profile, the imprint of
reflectivity variations is extremely mild and this approach fails most
of the times, especially when the observations are not critically
sampled in the spectral dimension. Instead, the inferred
reflectivities are usually dominated by a fringe pattern.
Therefore, a simplification to this scheme was proposed by
\citet{2013delacruz}, where reflectivities are assumed to be constant
across the FOV and the model can be simplified enormously as
convolutions become unnecessary and the mean spectrum can be shifted
using cubic Bezier interpolation. This is now the default in CRISPRED,
with the reflectivity fitting recommended only when the reflectivities
are actually going to be used for the analysis.

\subsection{Image restoration}
\label{sec:image-restoration}

\subsubsection{MOMFBD}
\label{sec:momfbd}

Images are restored from wavefront aberrations by means of the
Multi-Object Multi-Frame Blind Deconvolution \citep[MOMFBD;
][]{noort05solar} method, based on the algorithm by
\cite{lofdahl02multi-frame}. Not using any assumptions about the
aberration statistics, the method can compensate for optical
aberrations, both from the atmosphere and from the optics, and partial
compensation by the AO does not introduce any extra complexity.
However, we do take the aberration statistics into account by using
atmospheric Karhunen--Lo\`eve modes
\citep[KL,][]{roddier90atmospheric} in the expansion of the pupil
phase. These modes are constructed so they are both orthogonal on a
circular aperture and statistically uncorrelated assuming atmospheric
turbulence following a Kolmogorov model.

The correction is limited by the finite parametrization of the
wavefront, so high-order aberrations from both sources are not
corrected. The uncorrected aberrations show up in the restored data as
a stray light component. By default, we are currently using the 35
most significant atmospheric KL modes as basis functions for expanding
the unknown wavefronts. This number was set to about the same as the
number of modes corrected by the AO. However, as the AO was recently
upgraded to correcting 85 modes, the default number of KL modes
routinely used in MOMFBD will probably be increased.

The MOMFBD algorithm is based on the assumption that the aberrations
are isoplanatic, so that the point spread function (PSF) is constant
within the FOV. In order to make this assumption hold, at least
approximately, the processing is done on overlapping subfields of a
$\sim$5 arcseconds squared. This makes it possible to find the PSF and
deconvolve the raw data by use of fast Fourier transforms. In the
process, each subfield is low-pass filtered individually based on the
local signal-to-noise ratio (SNR). The restored subfields are then
mosaicked, resulting in a restored version of the entire FOV.

\subsubsection{Alignment}
\label{sec:alignment}

Physical quantities like magnetic field strengths and line-of-sight
velocities are encoded as small variations in the intensity. Stokes
maps and Dopplergrams are calculated as linear combinations of
intensity images from different modulator states or wavelength tunings
(here: just \emph{states}). Errors in alignment of the images involved
will show up as artifacts in the calculated quantities. However,
because the intensity structures differ from state to state, cross
correlation methods are not useful for alignment of images from
different states.

Given accurate information about the alignment of cameras from the
pinhole array (see Sect.~\ref{sec:pinhole-calibration}), the MOMFBD
joint processing of images from the WB camera with the variable states
images from two NB cameras will make all the restored images from a
single scan aligned to subpixel precision \citep{noort05solar}. The
reason this works is that all the WB images go into making a single
restored image (the \emph{anchor} image), forcing the estimated
wavefronts (mainly the tip and tilt components) to be such that they
align the raw WB images. Every NB image is simultaneous to one of the
WB images, and their estimated wavefronts are constrained to differ by
exactly the fixed tip and tilt offset that align the cameras.

In spite of the subfielding, there are residual anisoplanatic
effects in the restored data. Because the raw images for the different
NB states are not collected simultaneously, these residuals vary from
state to state. This results in relative geometric distortions on
scales smaller than the subfields. In the CRISPRED pipeline, we have
implemented the improvement suggested by
\citet{henriques12three-dimensional}, capable of removing also such
differential stretching from the restored NB images: In addition to
the restored WB image based on the entire sequence of raw WB images,
the MOMFBD program makes a set of additional restored WB images, each
based on only the raw images corresponding to the raw NB images of
each state. We then calculate alignment parameters (stretch vectors)
on a grid, that align this WB image with the anchor WB image, and
apply them to the corresponding restored NB image. Note that this does
not remove the variable blurring at scales smaller than the subfields.

\subsection{Prefilter correction}
\label{sec:prefilter}

Most FPI instruments use a prefilter to remove parasitic transmission
peaks that are present in the transmission profile. The observed
spectra are weighted by the profile of this prefilter, which must be
compensated for. In Sect.~\ref{sec:char} we estimate the spatially
averaged quiet-Sun profile from the flat-field data. Similarly to
\citet{2011schnerr}, we degrade the FTS solar atlas \citep{ftsatlas}
with a theoretical CRISP profile and a theoretical prefilter curve to
reproduce our observed quiet-Sun profile.

We adopt a Lorentzian profile to describe the prefilter, with the
following parameters:
\begin{equation}
  P(\lambda;p_1, p_2,p_3,p_4,p_5)
  = \frac{p_1}{\displaystyle 1+\bigg (2\cdot\frac{\lambda-p_2}{p_3}\bigg )^{2p_4}}
  \cdot (1+\lambda \cdot p_5),
\end{equation}
where $\lambda$ is the wavelength offset from line center, $p_1$ is a
scaling factor, $p_2$ is the central wavelength of the prefilter
relative to line center, $p_3$ is the FWHM of the prefilter,
$p_4$ is the number of cavities of the prefilter (normally fixed to 2)
and $p_5$ is a term to account for asymmetries in the prefilter curve.
Our implementation allows to fix the value of some parameters, which
can be useful if the number of observed line positions is small or
restricted to a small wavelength range.

The estimated prefilter curve is used to correct the intensities of
the science data (see Sect.~\ref{sec:time}).

\subsection{Post-processing of time-series}
\label{sec:time}

Image restoration is performed on each line scan individually. Forcing
the average of the tip-tilt corrections to be zero yields restored
images where the geometric distortions from anisoplanatism are largely
removed. However, some jitter may still be present, which becomes
apparent when combining many scans into time-series. For the same
reason, there can be residual alignment errors between scans. Finally,
the alt/az turret introduces image rotation along the day as the
relative angle of the mirrors changes within the telescope. In this
section we describe how we create homogeneous time-series.

Once the data are restored with MOMFBD and the polarimetric
calibration has been applied, we compute time dependent corrections to
obtain a homogeneous time series using the WB images. In a second
step, we apply those corrections to the NB images to create
science-ready data cubes. Polarimetric data sets result in sequential
cubes with shape $n_x \times n_y \times n_\lambda \times
n_\text{Stokes} \times n_t$, whereas non-polarimetric data have one
dimension less, $n_x \times n_y \times n_\lambda \times n_t$. The
output is compatible with CRISPEX\footnote{Now included in Solarsoft
  and available on the web at
  \url{http://folk.uio.no/gregal/crispex/}.}, a versatile tool for the
exploration of multi-dimensional data sets by \citet{2012vissers}.

Image field rotation angles are computed theoretically from the
position of the Sun on the sky and a telescope pointing model. After
de-rotation, the WB images are co-aligned using cross-correlation
methods. The residual rubber-sheet motions that usually appear among
the different time steps, are removed by resampling following
\citet{shine94high-resolution}. In short, the idea is to divide the
images in small subfields and compute the shift that aligns each
subfield with a reference. These corrections are also applied to the
cavity map, resulting in a different wavelength correction for each
time-step.

Variations in the intensities are expected as the elevation of the Sun
changes on the sky. We compute variations in the mean WB intensity
during the sequence and compensate for them. The data set is also
corrected for the prefilter curve that we measured in
Sect.~\ref{sec:prefilter}.

\section{Discussion}
\label{sec:discussion}

In this study, we present a new approach to compensate for the effect
of cavity-errors prior to image {restoration}. This method is
more accurate and, fundamentally more correct than the approximation
presented in \citet{2011schnerr}, and we advise to always use it to
process data from FPI instruments that are mounted in a telecentric
setup.

We also present improvements to the pinhole calibration process needed
to make the MOMFBD program output restored images from different
states that are well aligned, as well as the results of a new
procedure for measuring the image scale of image data from any
high-resolution solar telescope.

We have developed a processing pipeline for reducing data from the
SST/CRISP instrument in the form of an object-oriented framework. This
is constructed using existing routines and detailed knowledge of the
instrument, and produces science-ready data sets from the raw images.
However, routines that are specific to other instruments can be easily
incorporated in the future, i.e., the forthcoming CHROMIS instrument,
a Fabry--P\'erot interferometer to observe the \ion{Ca}{ii}~H
($\lambda3934$) and~K ($\lambda 3968$) lines. Note that many of the
routines included in this framework are also useful as building blocks
for building one-shot scripts for unusual data sets.

The transmission profiles of the FPI etalons were computed (see
Sect.~\ref{sec:crisp}) assuming perpendicular incidence of the light.
In reality, the beam is slowly converging and the low-resolution
etalon is tilted to reduce ghost images. Both the tilt angle and the
extreme angles from the beam convergence are $\arctan(1/165) \approx
0\fdg17$. These two effects in practice make the total CRISP profile
slightly wider and sightly asymmetric. For simplicity, we have adopted
this simplified description Eq.~(\ref{eq:eta}), but we plan to
investigate those effects in the future and include compensation for
them in the pipeline if necessary.

Some known effects are currently not included in the pipeline. The
finite area of the detector pixels causes a slight reduction of signal
at the highest detected spatial frequencies. The corresponding
modulation transfer function (MTF) also includes a wavelength
dependent component from charge diffusion that is more difficult to
model \citep{stevens94analytical}. Compensation for the geometric part
of the detector MTF would not be difficult to add but the
wavelength-dependent charge-diffusion effect requires either careful
calibrations or information from the manufacturer that we do not have.
There is ongoing work for the characterization of the various sources
of straylight at the SST and removing their effects
\citep{scharmer10high-order,lofdahl12sources}. When we have a good
enough model for this, a correction may be added to the pipeline.

The way flat-fields (flats) are made, by averaging many short-exposure
images collected while the telescope moves in small circles around the
center of the solar disk, allows the assumption that the spectrum
present at each pixel corresponds to a spatially averaged quiet-Sun
profile at disk center. The presence of active regions can sometimes
be inconvenient, forcing the observer to trace more elliptical
trajectories. To avoid distortions in the quiet-Sun profile, one could
normalize the NB flats by use of the mean intensity of the WB flat
images (acquired simultaneously with the NB images) as they should
average to the same value for all wavelengths. We are planning to
implement this as an optional step.
 
As part of the SOLARNET project, there is an effort to prepare for the
inclusion of data from ground-based telescopes in a virtual solar
observatory, similar to what is done for data from space-based solar
telescopes. Recommendations for file formats and meta data are
expected in the fall of 2014 and we aim to make CRISPRED compliant as
soon as possible.

\begin{acknowledgements}
  We thank Prof. G. Scharmer and Dr. D. Kiselman for illuminating
  discussions. The Swedish 1-m Solar Telescope is operated on the
  island of La Palma by the Institute for Solar Physics in the Spanish
  Observatorio del Roque de los Muchachos of the Instituto de
  Astrof{\'\i}sica de Canarias. This work was carried out as a part of
  the SOLARNET project, funded by the European Commission’s 7th
  Framework Programme under grant agreement no. 312495. We use
  subroutines from the IDL astronomy User's Library
  \citep{1993ASPC...52..246L}. IDL\textsuperscript{\textregistered} is
  a trademark of Exelis Inc. This research has made use of NASA's
  Astrophysics Data System (ADS).
\end{acknowledgements}

\balance
\bibliography{jaime,bib-strings_aa,ads,svst,mats,mats.lofdahl}


\appendix

\section{CRISP Prefilters}  
\label{ap:pref}

\begin{figure}[h]
  \includegraphics[width=1.0\hsize]{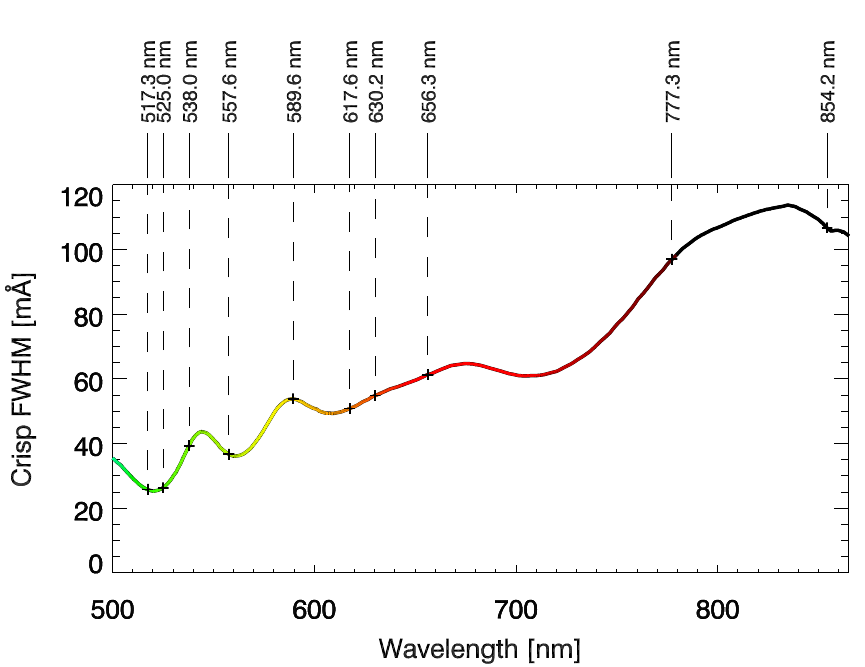}
  \caption{FWHM of the CRISP transmission profile as a function of
    wavelength. We have indicated those lines that are currently
    observable (i.e. a prefilter is available).}
  \label{fig:crispw}
\end{figure}

\begin{table}[h]
  \caption{List of prefilters for the CRISP instrument.}              
  \label{tab:pref}      
  \centering                                      
  \begin{tabular}{c  c  @{\quad}ll@{\quad}c c}          
    \hline\hline                        
    \noalign{\smallskip}                 
    CWL [nm] & FWHM [nm] & \multicolumn{3}{c}{Main diagnostic} & PC\\    
    \noalign{\smallskip}                 
    \hline                                   
    \noalign{\smallskip}                 
    517.33 & 0.30 && \ion{Mg}{i}&517.2\\
    525.01 & 0.30 && \ion{Fe}{i}&525.0 & \checkmark\\
    538.20 & 0.34 && \ion{C}{i}&538.0 & \checkmark\\
    557.80 & 0.30 && \ion{Fe}{i}&557.6\\
    \llap{$^*$}587.6\phantom{0} & 0.4\phantom{0} &&
    \ion{He}{i}&587.6 \\
    589.70 & 0.39 && \ion{Na}{i}&589.6\\
    617.39 & 0.42 && \ion{Fe}{i}&617.3 & \checkmark\\
    630.26 & 0.44 && \ion{Fe}{i}&630.2 & \checkmark\\
    656.38 & 0.49 && \ion{H}{i}&656.3\\
    777.24 & 0.77 && \ion{O}{i}&777.2 & \checkmark\\
    854.16 & 0.93 && \ion{Ca}{ii}&854.2 & \checkmark\\
    \noalign{\smallskip}                 
    \hline                                             
  \end{tabular}
  \tablefoot{$^*$Ordered from the manufacturer at the time of this
    publication. PC: Calibrated for polarization effects, see
    Sect.~\ref{sec:polcal}.} 
\end{table} 

\newpage 

Currently ten prefilters are available to observe selected spectral
lines, which have been indicated in Fig.~\ref{fig:crispw} and detailed
in Table~\ref{tab:pref}. Selection of the appropriate prefilter for
each observing sequence is done by means of a filter wheel that is
located near a focal plane (see Fig.~\ref{fig:opticalpath}). The
central wavelength (CWL) of the filters are slightly to the red of the
line core for which they are designed. The whole filter wheel sits on
a motorized tiltable mount, which allows tuning of the filters to the
exact wavelengths. In some cases this allows easy observation of other
interesting spectral lines close to the primary one.

\end{document}